\documentclass[a4paper,fleqn,usenatbib]{mnras}

\usepackage{newtxtext,newtxmath}

\usepackage[T1]{fontenc}
\usepackage{ae,aecompl,multirow}
\usepackage{hyperref}
\usepackage{ragged2e}
\usepackage{amsmath}

\makeatletter
\def\footnoterule{\kern-3\p@
  \hrule \@width 2in \kern 2.6\p@} 
\makeatother

\usepackage{graphicx}	
\usepackage{amsmath}	
\usepackage{amssymb}	
\usepackage{subfig}

\usepackage{caption}
\title[The synchrotron spectral nature of RGB\,J0710+591]{Unravelling the unusually curved X-ray spectrum of RGB\,J0710+591 using \emph{AstroSat} observations}

\author[P. Goswami et al.]
{Pranjupriya Goswami$^{1}$\thanks{E-mail: pranjupriya.g@gmail.com},
Atreyee Sinha$^{2}$,  
Sunil Chandra$^{3}$, 
Ranjeev Misra$^{4}$, \newauthor
Varsha Chitnis$^{5}$,
Rupjyoti Gogoi$^{1}$, 
Sunder Sahayanathan$^{6, 7}$,
C. S. Stalin$^{8}$,\newauthor 
K. P. Singh$^{9}$, and 
J. S. Yadav$^{10}$
\\ 
$^{1}$Department of Physics, Tezpur University, Napaam - 784028, India. \\ 
$^{2}$AstroParticule et Cosmologie, Universit\'e Paris Diderot, CNRS/IN2P3, CEA/Irfu, Observatoire de Paris, Sorbonne Paris Cit\'e, 10, \\
 rue Alice Domon et L\'eonie Duquet, 75205 Paris Cedex 13, France. \\ 
$^{3}$Centre for Space Research, North-West University, Potchefstroom, 2520, South Africa. \\
$^{4}$Inter-University Center for Astronomy and Astrophysics, Post Bag 4, Ganeshkhind, Pune - 411007, India.\\
$^{5}$Tata Institute of Fundamental Research, Homi Bhabha Road, Colaba, 400 005 Mumbai, India. \\
$^{6}$Astrophysical Sciences Division, Bhabha Atomic Research Centre, Mumbai - 400085, India. \\
$^{7}$Homi Bhabha National Institute, Mumbai 400094, India\\
$^{8}$Indian Institute of Astrophysics, Block II, Koramangala, Bangalore-560034, India. \\
$^{9}$Indian Institute of Science Education and Research, Mohali, Knowledge city, Sector 81, SAS Nagar, Manauli, Punjab - 140306. India. \\
$^{10}$Department of Physics, Indian Institute of Technology, Kanpur 208016, India. 
}
\date{}
\pubyear{2019}

\begin{document}
\label{firstpage}
\pagerange{\pageref{firstpage}--\pageref{lastpage}}
\maketitle

\begin{abstract}

We report the analysis of simultaneous multi-wavelength data of the high energy peaked blazar RGB\,J0710+591 from the LAXPC, SXT and UVIT instruments 
on-board \emph{AstroSat}. The wide band X-ray spectrum (0.35 -- 30 keV) is modelled as synchrotron emission from a non-thermal distribution of high energy 
electrons. The spectrum is unusually curved, with a curvature parameter $\beta_p \sim 6.4$ for a log parabola particle distribution, or a 
high energy spectral index $p_2 > 4.5$ for a broken power-law distribution. The spectrum shows more curvature than an earlier quasi-simultaneous analysis of 
\emph{Swift}-XRT/\emph{Nu}STAR data where the parameters were  $\beta_p \sim 2.2$ or $p_2 \sim 4$. It has long been known that a power-law electron 
distribution can be produced from a region where particles are accelerated under Fermi process and the radiative losses in acceleration site decide the maximum 
attainable Lorentz factor, $\gamma_{max}$. Consequently, this quantity decides the energy at which the spectrum curves steeply. We show that such a 
distribution provides a more natural explanation for the \emph{AstroSat} data as well as the earlier XRT/\emph{Nu}STAR observation, making this as the first 
well constrained determination of the photon energy corresponding to $\gamma_{max}$. This in turn provides an estimate of the acceleration time-scale as a 
function of magnetic field and Doppler factor. The UVIT observations are consistent with earlier optical/UV measurements and reconfirm that they plausibly 
correspond to a different radiative component than the one responsible for the X-ray emission.  

\end{abstract}



\section{Introduction}

Non-thermal emission observed from BL Lac class of AGNs are essentially dominated by emission from the relativistic jet which is directed towards the observer with a small angular separation. It is observationally featured by rapid variability from Doppler-boosted emission over a wide range of wavelengths from radio to $\gamma$-rays \citep{Blandford+1978, Urray+1995}. The spectral energy distribution (SED) of BL Lacs is double peaked, and this unfolds information on the various physical processes and origins to the non-thermal emission from the jet. The low energy component includes radio -- X-ray emission and is well understood to be synchrotron emission from a relativistic electron distribution \citep{Ghisellini+1989}. However, the origin of the high energy component from X-ray -- very high energy (VHE) $\gamma$-ray emission is still unclear. It is generally modelled as inverse Compton (IC) radiation by the same electron population accounted for the synchrotron emission, or external photon field from the broad line region (BLR) and the accretion disc (leptonic models, \cite{Bloom1996}). This may also due to hadronic interactions between accelerated protons $\&$ electron-positron pair or muon cascades (hadronic models, \cite{markus2010}).

The synchrotron peak in high energy peaked blazars (HBLs) typically lies in UV -- X-ray energies and the hard X-ray spectrum shows a steep spectra \citep{mislav2016, bartoli2016}. Besides this, the X-ray spectrum also shows smooth curvature around the peak and a mild curvature in the falling part for certain HBL sources. The X-ray curvature possibly indicates an energy dependence of the particle acceleration probability, which results in a log-parabola type particle distribution \citep{Massaro2004a}. Alternatively, an energy-dependent electron diffusion can also explain this curvature at hard X-ray energies \citep{pgoswami2018}. It is often observed that the synchrotron peak shifts towards higher energy X-ray energies during flaring episodes in the range from few eV to few tens of keV and is evident in many sources, e.g., Mkn421 \citep{Tramacere2009, Sinha2015_421},  Mkn 501 \citep{Pian1998}, and 1ES 2344+514 \citep{Giommi2000}. 

Interestingly, certain HBLs are known to have exceptionally high synchrotron peak even during their quiescent states. \cite{Costa2001} initiated an extensive study on 5 BL Lac type sources with BeppoSAX observations covering a wide range of energy 0.1-100 keV. For four sources the peaks were estimated at hard X-ray energy 1-5 keV and for 1ES\,1426+428, the synchrotron peak appeared at energy above 100 keV with a flat power law spectrum ($\alpha$ $<$ 1). Later, a few other sources 
are also observed to show this extreme synchrotron peak and consistently their Compton peak can reach up to 100 -200 GeV with hard very high energy (VHE) spectrum e.g,  RGB J0710+591 \citep{Acciari2010b}, 1ES 0347-121 \citep{Aharonian2007b}. These high energy peaked sources are often termed as extreme high energy peaked BL Lacs (EHBLs).

RGB\,J0710+519 (z=0.125) is an EHBL, first discovered by \emph{HEAO} A-1 and subsequently detected in VHE $\gamma$-rays with VERITAS array of atmospheric Cherenkov telescope during 2008 December and 2009 March \citep{Ong2009}. The preliminary studies on this source indicate the spectral hardening at TeV energies and the extreme Compton behaviour \citep{Nieppola2006, Abdo2009}. 
The spectral behaviour of this source has been studied by \cite{Acciari2010b} for synchrotron and Compton spectral components using VHE $\gamma$-ray observation by VERITAS, supplemented with the multi-wavelength observations from \emph{Fermi} and \emph{Swift}. The time averaged \emph{Swift}-XRT spectra during 20 February - 2 March, 2009 can be explained by an absorbed power-law model with photon index $\sim$ 1.86. This spectral hardening in X-ray spectrum is the indication that the synchrotron peak can reach up to 10 keV or above. A comprehensive study on this object and a few other extreme TeV BL Lacs involving more recent observations has been carried out by \cite{Costa2018}. The authors discussed the X-ray spectrum using the simultaneous \emph{Swift} and \emph{Nu}STAR observations and, interestingly, the synchrotron peak for this source was constrained for the first time at energy $\sim$ 3.5 keV. In addition, the authors modelled the complete SED with various multi-wavelength data from \emph{Fermi} satellite and other available data in the energy range extending from radio to VHE $\gamma$-rays. The optical/UV emission of this source is clearly not an extrapolation of the X-ray spectrum and a possible interpretation could be that different regions are responsible for X-ray and UV/optical emissions \citep{Acciari2010b, Costa2018}.  

In this work, we perform a detailed investigation of the extreme synchrotron behaviour of the BL Lac source RGB\,J0710+591, and establish the nature of X-ray spectrum using strictly simultaneous multi-waveband \emph{AstroSat} data for the first time. Our aim is to constrain the synchrotron peak within the limit of observed X-ray energies and study the particle acceleration mechanisms responsible for the curved synchrotron spectrum. The \emph{AstroSat} data is supplemented with the simultaneous \emph{Swift}-XRT and \emph{Nu}STAR observations in energy range 0.3 -- 79 keV to compare the spectral transitions during different flux states. Furthermore, we discuss the UV and optical emissions observed by the UV/optical instruments of \emph{AstroSat} and \emph{Swift}. The observations and data reduction procedures are described in section \ref{astrosat} and \ref{swnu}. In section \ref{analysis}, we perform spectral analysis. The interpretation of the results is discussed in Section \ref{discussion}.      

\section{AstroSat observations}
\label{astrosat}

\emph{AstroSat}, India's first multi-wavelength space observatory launched in September 2015, has five scientific instruments onboard covering a wide range of energies from UV to hard X-ray \citep{Agarwal2006, KP2014, Rao2016}. The instruments onboard \emph{AstroSat} are: Soft X-ray \linebreak focusing Telescope (SXT), Large Area X-ray Proportional Counters (LAXPC), UltraViolet Imaging Telescope (UVIT), Cadmium Zinc Telluride Imager (CZTI) and Scanning Sky Monitor (SSM). RGB\,J0710+591 was observed by SXT (as primary instrument), LAXPC and UVIT on 19 Novemeber, 2016 for 1 pointing (Obs ID: A02$\_$085T02$\_$9000000808; \linebreak Table \ref{observation}). The observations and  data reduction techniques are discussed in the following sections  \ref{sxt}, \ref{lxpc} and \ref{uvitinstru}.      

\subsection{SXT}
\label{sxt}

The SXT is a focusing telescope capable of X-ray imaging and spectroscopy in the energy range 0.3 -- 8.0 keV with 2$'$ angular resolution and FOV of $\sim$ 40$'$ diameter \citep{KP2016, KP2017}. The Level-1 SXT data observed in the photon counter (PC) mode were first processed with {\tt sxtpipeline} available in the SXT software (\emph{AS1SXTLevel2}, version 1.4b). The pipeline calibrates the source events and extracts Level-2 cleaned event files for the individual \linebreak orbits. The cleaned event files of all the 10 orbits are then merged into a single cleaned event file using {\tt SXTEVTMERGER} tool developed by the instrument team to avoid the time-overlapping events from the consecutive orbits. The {\tt XSELECT} (V2.4d) package built-in \emph{HEAsoft} is used to extract the source spectrum from the processed Level-2 cleaned event files. We selected the source region as a circular region of 14 arcmin radius centred at the source position, which encompasses more than 90$\%$ of the source pixels. The background spectrum {\tt $"$SkyBkg$\_$comb$\_$EL3p5$\_$Cl$\_$Rd16p0$\_$v01.pha$"$}, a composite product using a deep blank sky observation, distributed by the instrument team is used for spectral analysis. We have made use of the ancillary response file (ARF) {\tt $"$sxt$\_$pc$\_$excl00$\_$v04$\_$20190608.arf$"$} (version 4.0) released recently by the instrument team. The used response file ({\tt $"$sxt$\_$pc$\_$mat$\_$g0to12.rmf$"$}), ARF and background are available at the SXT POC website \footnote{\href{www.tifr.res.in/~astrosat$\_$sxt}{www.tifr.res.in/~astrosat$\_$sxt}}. The source spectrum was then grouped using the {\tt grppha} tool to ensure a minimum of 60 counts per bin. We obtained the net count rate for the SXT spectrum with 0.26 ct/s in the enrgy range 0.3 -- 7.0 keV.

\subsection{LAXPC}
\label{lxpc}

The LAXPC is one of the primary instruments on board \emph{AstroSat} and consists of three identical co-aligned X-ray proportional counter units providing with high time resolution ($\sim$ 10 $\mu$s) covering 3 -- 80 keV energy band \citep{Yadav2016, Antia2017, Agarwal2017LXPC, Misra2017}. The propotional counter units are named as LAXPC10, LAXPC20 and LAXPC30 with each detector having an effective area of $\sim$ 2000 cm$^2$. To process the Level-1 LAXPC data, the \emph{laxpc$\_$soft} packages were used which is based on Fortran codes developed by the instrument team, available at the \emph{AstroSat} Science Support Cell (ASSC) website \footnote{\href{http://astrosat-ssc.iucaa.in/?q=sxtData}{http://astrosat-ssc.iucaa.in/?q=sxtData}}. The data reduction procedures involve the generation of event files, standard GTI files of good time intervals to avoid Earth occulation and the South Atlantic Anomaly, and finally, extraction of source spectrum. To perform these, we used tools {\tt laxpc$\_$make$\_$event} and {\tt laxpc$\_$make$\_$stdgti} which are in-built in the \emph {laxpc$\_$soft} package. Data from source free sky regions observed within a few days of the source observation are used to generate and model the background and appropriate scaling is performed depending on the orbit. Finally, to generate the lightcurve and spectra, {\tt laxpc$\_$make$\_$lightcurve} and {\tt laxpc$\_$make$\_$spectra} tools were used. For the faint sources like AGNs, the background estimation is not straightforward, as the background starts dominating over the source counts. Therefore, the background was estimated from 50--80 keV counts where background seems relatively steady. The data of each LAXPC unit (LAXPC10 and 20) is reduced separately and data from the top layers (layer 1) from each units are recommended to use for faint sources. LAXPC30 data were not considered due to the continous gain shift observed in this unit, suspected to be caused by a gas leakage \citep{Antia2017}, while the data from LAXPC10 unit were unstable. Thus, in this analysis, we have only used the LAXPC data from LAXPC20. Using a total of 67.3 ks of useful data from the top layer of LAXPC20 (LX20-L1) the total count rate was estimated to be 0.5 ct/s in the enrgy range 3.0--30 keV.

\subsection{UVIT}
\label{uvitinstru}

UVIT is primarily an imaging telescope on board \emph{AstroSat} and consists of 3 channels in visible and ultraviolet(UV): VIS (320-550 nm), NUV (200-300 nm) and FUV (130-180 nm) wavelengths provided with absolute spatial resolution (FWHM $<$ 1.8$"$) images in a field of view $\sim$ 28$'$ \citep{kumar2012, Annapurni2016, Tandon2017}. The detail information on the detectors and the UVIT filetrs are available at the \emph{AstroSat} UVIT website \footnote{\href{http://uvit.iiap.res.in/}{http://uvit.iiap.res.in/}}. UVIT Level-1 data were processed with \emph{UVIT Level-2 Pipeline} (Version 5.6), software accessible at ASSC website and data reduction procedures were followed as recommended in the manual provided with the software packages. The pipeline generates the full frame astrometry fits images for each filter in NUV and FUV channels seperately which are corrected for flat-fielding and drift due to rotation. To extract the count from the fits images corresponding to each filter, aperture photometry was performed using \emph{IRAF} (Image Reduction and Analysis Facility) software tool. An aperture of 50 pixels radius size was selected to do photometry which encompasses $\sim$ 98$\%$ of the source pixels. The extracted counts were then converted into fluxes for each filter using the unit conversion as suggested by \cite{Tandon2017}. The fluxes were then corrected for Galactic interstellar extinction \citep{Fitzpatrick} with the values of A$_B$ = 0.139 and E$_{(B-V)}$ = 0.034 taken from NED. We have used the UVIT data for 5 filters which contain 3 NUV (NUVB13, NUVB4 and NUVN2) and 2 FUV filters (BaF2 and Silica) (See Table \ref{uvit}).

\begin{table*}
\centering
\caption{Details of the observations from various instruments of \emph{AstroSat}, \emph{Swift} and \emph{Nu}STAR missions}
\vspace*{0.1cm}
\begin{tabular}{cccccc}
\hline \hline  
 
Instrument& Observation ID & Observation date & Exposure &  \\
 &   & (dd-mm-yyyy) & (ks)\\
\hline 
\textbf{SXT}&A02$\_$085T02$\_$9000000808 & 19-11-2016 T00:59:28& 20.25& \\
\hline 
\textbf{LAXPC-20}&A02$\_$085T02$\_$9000000808 &19-11-2016 T21:23:08 & 27.5& \\
\hline 
& & & XRT & UVOT \\
\textbf{Swift} & 00081693002 &02-09-2015 T00:03:16 & 4.43&4.49 \\
\hline 
\textbf{NuSTAR}&60101037004  & 01-09-2015 T12:11:08 & 26.4 & \\
\hline \hline

\end{tabular}
\label{observation}
\end{table*}

\begin{table*}
\centering
\caption{\emph{AstroSat}-UVIT filter details and the measured flux:}
\vspace{0.1cm}
\begin{tabular}{cccccc}
\hline \hline  

 Filter Name (slot) &   $\lambda_0$ ($\AA$) &  $\Delta\lambda$ ($\AA$) & Exposure (ks)& Flux (10$^{-12}$  erg cm$^{-2}$ s$^{-1}$)\\
\hline  
\textbf{FUV}& & & &\\
BaF2 (F2)&1541 &380 & 5.35 & 2.36 $\pm$ 0.02 \\
Silica (F5)&1717 &125 & 6 & 2.26 $\pm$ 0.03 \\
\hline
\textbf{NUV} & & & \\
NUVB13 (F3)&2447 & 280& 3.8 & 2.01 $\pm$ 0.07 \\
NUVB4 (F5)& 1632 &275 & 8.7 & 2.02 $\pm$ 0.18 \\
NUVN2 (F6)& 2792 & 90& 6.6 & 2.00 $\pm$ 0.16 \\

\hline \hline

\end{tabular} \\
Note: $\lambda_0$ and $\Delta\lambda$ are the central wavelength and the effective band width of the filters. \\ The given fluxes are corrected for the Galactic extinction.
\label{uvit}
\end{table*}

\section{\emph{Swift} and \emph{Nu}STAR observations} 
\label{swnu}

To compare the \emph{AtroSat} results and study the evolution of X-ray spectrum during different flux states, various instrument data are taken into account. The source has been monitored by \emph{Swift} and \emph{Nu}STAR on a few occasions. The two \emph{Swift} pointing (Obs ID: 00031356055, 00031356056) are contemporaneous with the \emph{AstroSat} observations within a gap of $\sim$10 days, however, these observations are discarded due to low count rates and nonavailability of simultaneous \emph{Nu}STAR observations. Hence, for the completeness of this work we have repeated the analysis from previous studies \cite{Costa2018} with \emph{Swift}-XRT and \emph{Nu}STAR observations from 2015. The details of the observations are reported in Table \ref{observation}.

We collected \emph{Swift} and \emph{Nu}STAR archival data that are available in NASA's HEASARC interface \footnote{\href{https://heasarc.gsfc.nasa.gov/}{https://heasarc.gsfc.nasa.gov/}}. The standard data reduction procedures were followed to analyse the data from various instruments UVOT, XRT in \emph{Swift} and \emph{Nu}STAR. \emph{Swift}-XRT \citep{Burrows2005} data were processed with the \emph{XRTDAS} software package (Version 3.0.0) built-in \emph{HEAsoft} (Version 6.22). The source was observed in photon counting (PC) mode and corrected for pile-up by excluding central 3-pixels in the source region as the initial source count rate was above 0.5 ct/s. \emph{Swift}-UVOT \citep{Roming2005} observations of 6 UV filters: U, V, B, UVW1, UVW2 and UVM2 in UV and optical wavelengths \citep{Poole2008}, were included in this study. The data from both the telescope FPMA and FPMB in \emph{Nu}STAR \citep{Harrison2013} were processed using the \emph{NuSTARDAS} software  package (Version 1.4.1). The details of stepwise data reduction procedure for \emph{Swift}-XRT and \emph{Nu}STAR data are discussed in \cite{pgoswami2018}.

\section{Spectral Models and analysis}
\label{analysis}

Previous studies confirm spectral breaks with a smooth curvature in the X-ray spectrum of this source and therefore, a simple power-law interpretation is not adequate. A broken power-law or a log-parabola distribution can possibly explain this. Several authors have reported the curvature property for the combined \emph{Swift}-XRT and \emph{Nu}STAR observations using a log-parabola photon distribution \citep{Acciari2010b, Costa2018, Pandey2018}. The main focus of this work is to present the results from various X-ray instruments onboard \emph{AstroSat} observed in 2016. This can give us an idea about the spectral variations of the source during various flux states. 

\begin{figure}
\centering
	\includegraphics[scale=0.3,angle=270]{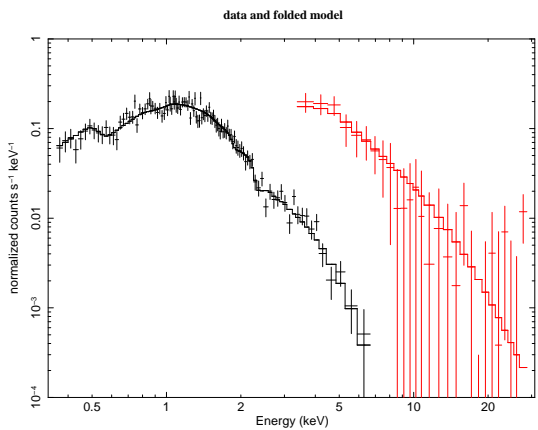}		
	\caption{Figure shows combined SXT-LAXPC20 data (SXT in black and LAXPC in red) fitted with a log-parabola photon spectrum model.}
	\label{plld}
\end{figure}

The spectral fittings were performed for each observation from various instruments using {\tt XSPEC} (Version 12.9.1) software package  distributed with \emph{HEASoft} \citep{Arnaud1996}. The correction of galactic absorption was done by using \emph{TBabs} model in {\tt XSPEC}, considering the value of equivalent-hydrogen column density (N$_H$) fixed at 4.44 $\times$ 10$^{20}$ cm$^{-2}$ and was kept fixed throughout the analysis. This value was estimated by online tool \footnote{\href{https://heasarc.gsfc.nasa.gov/cgi-bin/Tools/w3nh/w3nh.pl}{https://heasarc.gsfc.nasa.gov/cgi-bin/Tools/w3nh/w3nh.pl}} developed by the LAB survey group \citep{Kalberla2005}. A best-fit nominal gain offset of 0.03 keV as determined using {\tt gain fit} option with a fixed gain slope of 1 is used, as recommended by SXT instrument team. This significantly improves the fit statistics. Once best-fit gain parameters are decided, we have fixed these throughout the spectral fitting in order to save computation time while calculating the error bars and contours. To determine the relative cross-calibration uncertainties between two instruments, a multiplicative constant factor is used along with models and reported in Table \ref {logpar}. We use the X-ray observations corresponding to joint XRT (ID: 00081693002) -- \emph{Nu}STAR in the energy range 0.35 -- 50 keV and SXT -- LAXPC in the range 0.35 -- 30 keV for spectral fitting. Table \ref{logpar} shows the X-ray spectral fitting results of these data sets using a log-parabola function in-built in {\tt XSPEC}. The spectral fit and the residuals of the SXT-LAXPC data are shown in Figure \ref{plld}. Both the spectra show clear curvatures, and the SXT observation alone can accommodate the synchrotron peak with a sharp spectral turnover. We observe a slight change in the 2-10 keV average flux which in contrast, shows a considerable change in the curvature value. The synchrotron peaks of both the spectra are estimated by log-parabola model and well constrained within the observed X-ray energy range. Interestingly, the peak shifts by a significant fraction (by a factor of $\sim 3.5$) during these epochs. In addition, we also observe a strong curvature in the X-ray spectrum of AstroSat-SXT, which is changed by 1.2 - 1.7 times from the previous XRT-NuSTAR spectrum. Hence, the spectral slope estimated at 1keV and 10keV using the best-fit log-parabola model parameters changes sharply in case of SXT-LAXPC spectrum ($>1.5$) compared to XRT-NuSTAR spectrum ($\sim 1.07$). This spectral steepening of SXT-LAXPC spectra is also consistent with the low energy spectral peak. The spectral fitting was performed by keeping the cross-normalization parameter free for the spectra 
corresponding to XRT and SXT and fixed at 1.0 for LAXPC-20 and \emph{Nu}STAR. However, we observe that the cross-normalization between XRT and 
\emph{Nu}STAR is too low and the difference is of the order of $\sim$40$\%$. This is possibly due to the poor statistics of XRT data beyond 3 keV. Therefore, 
to keep this cross-normalization uncertainty under acceptable limit (typically $\lesssim 10-15 \%$, \citet{Madsen_2016}), we freeze this value at 0.85 for XRT. 
The change in $\chi^2$ for other values of cross-normalization within 20$\%$-15$\%$ for XRT observation are observed as 432.13 (429) for cross-normalization
 0.80, 471.13 for 0.85, 486.27 for 0.90 and 518.57 for 0.95. Fixing the relative normalization causes marginal changes in curvature. Hence, for better 
understanding, we have reported the best-fit parameters for both the cases with the cross-normalization free and fixed at 0.85 for XRT. 

\begin{table*} 
\centering
\footnotesize
\caption{Best-fit parameters using log-parabola photon spectrum model} 
\vspace*{0.2cm}

\begin{tabular}{lcccccccr}
\hline \hline  

Observation  &  Constant * & $\alpha$    & $\beta$     & $\chi^2$ (dof) & E$_{syn,peak}$ &F$_{0.3-2\rm\,keV}$ & F$_{2-10\rm\,keV}$\\
&&&&& (keV) & (10$^{-12}$ erg cm$^{-2}$ s$^{-1}$) & (10$^{-12}$ erg cm$^{-2}$ s$^{-1}$) \\
\hline
SXT -- LAXPC-20    &  1.09 & 1.82 $^{+0.07}_{-0.07}$ & 0.64 $^{+0.18}_{-0.17}$ &111.15 (135)& 1.37 $^{+0.16}_{-0.13}$ & 8.53 $^{+0.12}_{-0.07}$  & 7.03 $^{+0.53}_{-0.44}$ \\
XRT -- \emph{Nu}STAR& 0.59    & 1.58 $^{+0.07}_{-0.07}$ & 0.38 $^{+0.04}_{-0.05}$ & 394.81 (428)& 3.63 $^{+0.34}_{-0.35}$ & 6.79 $^{+0.10}_{-0.17}$  & 10.01 $^{+0.06}_{-0.07}$ \\
                    &  0.85    & 1.28 $^{+0.05}_{-0.05}$ & 0.51 $^{+0.04}_{-0.03}$ & 457.31 (429)& 5.01 $^{+0.21}_{-0.22}$ & 6.78 $^{+0.16}_{-0.14}$  & 13.51 $^{+0.13}_{-0.12}$ \\

\hline \hline \\

\end{tabular} \\
\label{logpar}

\justifying \textbf{Note:} * The relative cross-normalization constant between two different X-ray instruments. The best-fit values for combined XRT-\emph{Nu}STAR spectrum are reported here correspond to the cross-normalization constant for SXT free and fixed at 0.85, while fixed at 1.0 for \emph{Nu}STAR . This parameter was kept at 1.0 for LAXPC and free for SXT observations in SXT-LAXPC20 spectrum. The errors are estimated within 90$\%$ confidence range based on the criterion used in {\tt XSPEC}.  
\end{table*}

It is clear from our spectral fitting results that the two X-ray spectra observed by XRT-\emph{Nu}STAR and SXT-LAXPC are significantly curved. The SXT-LAXPC spectrum appears to be more curved with steeper X-ray spectral index than XRT-\emph{Nu}STAR spectrum when fitted with a log-parabola model. For a better understanding about the emitting particle distribution and the intrinsic curvature, we reproduce the synchrotron spectrum assuming various underlying electron distributions. 

The synchrotron emissivity due to an electron distribution losing its energy in a magnetic field B is given by, \citep{Rybicki_Lightman}, 

\begin{align}\label{eq:syn_emiss}
	j_{\rm syn}(\nu)= \frac{1}{4\pi}\int P_{\rm syn}(\gamma,\nu)\,N (\gamma)\,d\gamma
\end{align}
Here, N($\gamma$) is the electron number density at dimensionless energy $\gamma$ and $P_{\rm syn}$ as the single particle synchrotron emissivity. Assuming a spherical emission region of volume V, the observed synchrotron flux F$_{\rm syn}$($\nu$) after accounting for relativistic beaming and cosmological evolution will be, \citep{Begelman1984},
\begin{align}\label{eq:obs_flux}
	F_{\rm syn}(\nu)= \frac{\delta_D^3(1+z)}{d_L^2} V 
	j_{\rm syn}\left(\frac{1+z}{\delta_D}\nu\right)\quad\quad\rm erg\,cm^{-2}\,s^{-1}\,Hz^{-1},
\end{align}
Here, z is the redshift of the source, d$_L$ is the luminosity distance and $\delta_D$ is the relativistic Doppler factor. The observed synchrotron spectral shape is determined by N($\gamma$) and hence different particle distribution or acceleration mechanism could possibly explain the observed X-ray spectral nature of this source in different flux states.

 To interpret the X-ray spectral curvature, we refit the \emph{AstroSat} and XRT-\emph{Nu}STAR observations as synchrotron spectrum due to different electron distributions and adding it as a local model in {\tt XSPEC}. We first assume a broken power-law distribution for N($\gamma$) given by, 
\begin{equation}\label{eq:bknpo_eq}
N(\gamma) d\gamma =
\begin{cases}
 \text{K $\gamma^{-p1}\,d\gamma, \quad\rm \gamma_{min} < \gamma < \gamma_b$}\\
 \text{K $\gamma_b^{(p2-p1)}\,\gamma^{-p2}\,d\gamma, \quad\rm \gamma_b < \gamma < \gamma_{max}$}\\

\end{cases}
\end{equation}
where, $\gamma_b$ is the break energy and p1 and p2 are two particle indices at low and higher energies. A broken power-law distribution is capable to fit both the spectra and the best-fit parameters are reported in Table \ref{xcomb}. The parameter E$_b$ is the break energy in the photon spectrum associated with the particle break energy $\gamma_b$. However, although the model shows reasonably good $\chi^2$ fit statistics, the model is inappropriate to explain the SXT spectrum as it demands higher curvature. The model fit parameters are not constrained, showing a large interval in the error estimation (p2$>$4.5). 

In order to explain the higher curvature, we assume for N($\gamma$) a log-parabola particle distribution which is given by 
\begin{align}\label{eq:lp_eq}
N(\gamma) d\gamma = K \left(\frac{\gamma}{\gamma_0}\right)^{-\alpha_p-\beta_p\,\rm{log}(\gamma/\gamma_0)} d\gamma
\end{align}
Here, $\alpha_p$ the particle spectral index at energy $\gamma_0$ and $\beta_p$, representing the curvature are the free parameters. K is the normalization constant. A log-parabola particle distribution model is statistically a better model than a broken power-law distribution with since it involves a lesser number of parameters (see Table \ref{xcomb}). The model parameters are better constrained in XRT-\emph{Nu}STAR spectrum, but showing larger interval for SXT-LAXPC spectrum while estimating errors. The estimation of the error for each parameter in all the models is within 90$\%$ confidence range. 

These results point out the exceptional curved feature of the SXT-LAXPC spectrum with unconstrained high energy index p2$>$4.5 in broken power-law and a large value of spectral curvature $\beta_p$ $\sim$ 6.4 in log-parabola particle distributions. In spite of good $\chi^{2}$ fit estimations, these models do not have a reasonable explanation to validate the unusual curvature in SXT-LAXPC spectrum. 
Further, this is consistent with the sharp spectral change witnessed in the photon spectra (Table \ref{logpar}); however, the change in the slope is more prominent in case of particle distribution. For XRT-\emph{Nu}STAR spectrum the change in the slope of the particle distribution corresponding to emission at 1keV and 10keV
is $\sim$ 3.9. On the other hand, this slope change in case of SXT-LAXPC spectrum is $\sim$ 8, which is almost twice that of 
XRT-\emph{Nu}STAR. This ensures the curvature in \emph{AstroSat} X-ray spectrum is considerably higher than that of XRT-\emph{Nu}STAR.

Alternatively, a steep spectral curvature can be an outcome of rapidly decaying particle distribution near the maximum available electron energy $\gamma_{max}$. To model this, we consider a scenario where the electrons are accelerated through Fermi acceleration process at the vicinity of a shock (Acceleration region, AR) and undergo synchrotron energy losses. The electron distribution N($\gamma$) in the AR will be governed by \citep{Kardashev1962},

\begin{align}\label{eq:AR}
	\frac{d}{d\gamma} \left[\left(\frac{\gamma}{t_a}-A\gamma^2\right)N(\gamma)\right] + \frac{N (\gamma)}{t_e}
         = Q_o \delta(\gamma-\gamma_0)
\end{align}
Where, t$_a$ and t$_e$ are the acceleration and the escape time scales, A$\gamma^2$ decides the radiative loss term and for simplicity we consider a monoenergetic electron injection with energy $\gamma_0$ into AR. The steady state solution in AR for constant t$_a$ and t$_e$ will then be \citep{kirk1998},

\begin{align}\label{eq:spectral_form}
N(\gamma) d\gamma = K \left(\frac{\gamma}{\gamma_{max}}\right)^{-p} \left(1-\frac{\gamma}{\gamma_{max}}\right)^{(p-2)} d\gamma
\end{align}
Here, $p=\left(1+\frac{t_a}{t_e}\right)$ is the particle spectral index. The maxium energy of the electron attained in AR will be decided by the rate of acceleration and radiative loss, $\gamma_{max} = \frac{1}{A t_a}$. The radiative loss term A, in equation \ref{eq:AR} is associated with the intrinsic magnetic field and given by, 
\begin{align}\label{eq:A}
A= \frac{4}{3}\frac{\sigma_T}{m_e c} \left(\frac{B^2}{8\pi}\right)
\end{align}
Where, $\sigma_T$ the Thomson cross-section. We convolve the number density given by equation \ref{eq:spectral_form} with the single particle emissivity to obtain the synchrotron spectrum which is incorporated as a local model in {\tt XSPEC}. The particle spectral index, p and the synchrotron photon energy E$_{max}$ corresponding to $\gamma_{max}$ are chosen as model parameters.

\begin{figure*}
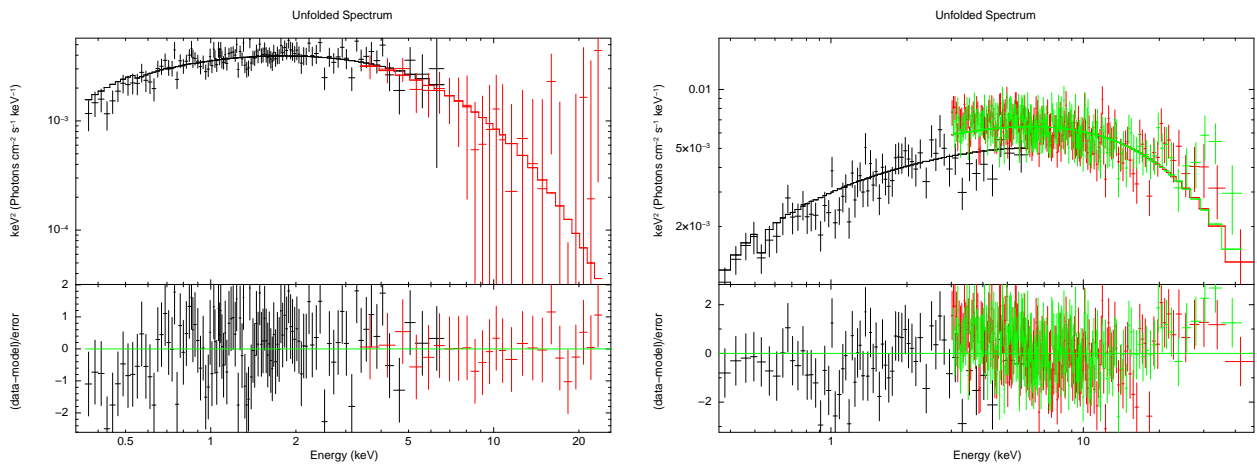

\centering
	\includegraphics[scale=0.33,angle=270]{astrosat_xray}
	\includegraphics[scale=0.33,angle=270]{swnu_updated}
		
	\caption{Figures show spectral fitting for combined SXT-LAXPC20 observations on LHS (SXT in black and LAXPC in red) and the simultaneous XRT and NuSTAR observation on RHS (Constant=0.85, XRT in black and the NuSTAR data from FPMA and FPMB are in red and green) using PL with $\gamma_{max}$ model.}
	\label{xray_fit}
\end{figure*}

The model shows satisfactory fits and the best-fit values of model parameters are tabulated in Table \ref{xcomb}. Figure \ref{xray_fit} shows the model spectral fit and the corresponding fit residuals for SXT-LAXPC (LHS) and XRT-\emph{Nu}STAR (RHS) observations. The parameters are well constrained within a small interval. This is indeed a remarkable result that the change in X-ray shapes can be interpreted in terms of a maximum Lorentz factor $\gamma_{max}$ in the acceleration zone that varies significantly over time. This model provides a more natural explanation to the observed spectral evolution and particularly the sharp curvature seen by SXT. The photon energy corresponds to $\gamma_{max}$ for both the spectra are well confined and can be estimated within the observation range.


\begin{table*} 
\centering
\scriptsize
\caption{Best-fit parameters using synchrotron spectrum models with broken power-law, log-parabola and PL with $\gamma_{max}$ particle distributions. The best-fit parameters correspond to the XRT-\emph{Nu}STAR spectrum are reported for both the cases with XRT cross-normalization free and fixed at 0.85.}
\vspace*{0.2cm}
\begin{tabular}{lccccccccccr}
\hline \hline  
Observation&\multicolumn{4}{c}{SXT -- LAXPC-20} & & \multicolumn{4}{c}{XRT -- \emph{Nu}STAR }& \\
\hline 

Broken  &  p1    &  p2 & E$_b$ (keV)    & $\chi^2$ (dof) && Const. &p1 & p2 & E$_b$ (keV) & $\chi^2$ (dof)\\
\cline{2-5}
\cline{7-11} 
Power-law & $<$ 2.02  & $>$ 4.5 &  2.44 $^{+0.44}_{-1.28}$  & 110.42 (134) &&0.59 & 1.49 $^{+0.37}_{-0.70}$  & 3.88 $^{+0.24}_{-0.16}$  & 2.77 $^{+0.79}_{-0.76}$  & 390.57 (427)   \\

&  & &     &  && 0.85 & $<$0.59 & 4.60 $^{+0.51}_{-0.23}$  &3.25 $^{+1.30}_{-0.20}$  &  463.82 (428) \\

\hline

Log-parabola  &  $\alpha_p$    & $\beta_p$     &  $\chi^2$ (dof) & && Const. & $\alpha_p$ &  $\beta_p$ & $\chi^2$ (dof)  \\
\cline{2-4}
\cline{7-10} 
 & 1.96 $^{+0.38}_{-0.61}$  & 6.41 $^{+4.35}_{-2.51}$ & 110.35 (135) & && 0.59 & 1.73 $^{+0.25}_{-0.29}$ & 2.17 $^{+0.40}_{-0.35}$ & 395.37 (428)&  \\
&  & &     &   && 0.85 &  0.40 $^{+0.29}_{-0.34}$ & 3.67 $^{+0.51}_{-0.45}$ & 471.13 (429)&  \\

\hline

PL with $\gamma_{max}$  &  p &  E$_{max}$ (keV)    & $\chi^2$ (dof) & && Const. & p &  E$_{max}$ (keV) & $\chi^2$ (dof)&  \\
\cline{2-4}
\cline{7-10} 
 &2.01 $^{+0.21}_{-0.01}$ & 10.36 $^{+0.98}_{-0.72}$ & 111.74 (135)&  && 0.59 & 2.25 $^{+0.11}_{-0.10}$ & 65.12 $^{+1.52}_{-1.24}$  & 423.00 (428) & \\
&  & &     &  && 0.85 & 2.11 $^{+0.11}_{-0.11}$ & 54.76 $^{+0.96}_{-0.86}$  & 440.96 (429)& \\

\hline \hline
\end{tabular} \\
\label{xcomb}
\end{table*}


\begin{figure*}
\centering
	\includegraphics[scale=0.5,angle=0]{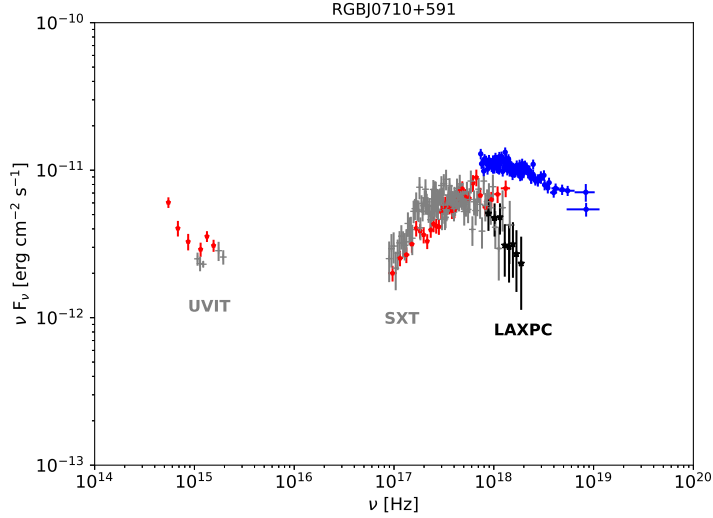}
		
	\caption{Figure shows the comparison between the SEDs with \emph{AstroSat} (SXT,UVIT: grey, LAXPC: black) and combined \emph{Swift} (XRT,UVOT: red) and \emph{Nu}STAR (blue) observations from November 2016 and September 2015. The SEDs are clearly different in shape, showing more curvature in SXT spectrum. The change in UV fluxes is marginal.}
	\label{sed_2}
\end{figure*} 

 
The best fit E$_{max}$ let us to estimate the maximum available electron energy in AR, with the knowledge of source magnetic field and Doppler factor
\begin{align}\label{eq:gmax}
\gamma_{max} = 4.02 \times 10^6 \left(\frac{\delta_D}{30}\right)^{-1/2}\rm\left(\frac{B}{0.011}\right)^{-1/2} \rm \left(\frac{E_{max}}{54.76}\right)^{1/2} 
\end{align}
Here, the choice of B and $\delta_D$ is chosen from \cite{Costa2018}. We estimate the $\gamma_{max}$ for XRT-\emph{Nu}STAR spectrum as 4.02 $\times$ 10$^6$ using the best-fit parameter E$_{max}$=54.76 (keV) for XRT cross-normalization 0.85. The estimated $\gamma_{max}$ changes by a factor 1.08, when we consider the XRT cross-normalization as 0.59. Since $\gamma_{max}$ is decided by the acceleration and radiative cooling timescales, using equation \ref{eq:A} and \ref{eq:gmax}, t$_a$ will be,
\begin{align}\label{eq:ta}
t_a = 1.59 \times 10^{6} \left(\frac{\delta_D}{30}\right)^{1/2}\rm\left(\frac{B}{0.011}\right)^{-3/2} \rm \left(\frac{E_{max}}{54.76}\right)^{-1/2}   secs
\end{align}

The estimation of $\gamma_{max}$ and t$_a$ from above equations demands prior knowledge of intrinsic magnetic field and Doppler factor. Unfortunately, this information can not be attained only from the \emph{AstroSat} observation and demand information at $\gamma$-ray energies. In reality, we expect to see significant changes in the magnetic field for different epochs, and hence the acceleration time scale. 
 Thus, we conclude that the observed spectral transition is associated with the variation in acceleration time scale or the magnetic field/Doppler factor over time.

Further, the \emph{AstroSat} UVIT observation in the near-far UV energy ranges (4-8 eV) shows an irregular shape and is clearly not an extrapolation of the X-ray spectrum. Figure \ref{sed_2} shows the broad optical/UV to X-ray SEDs of \emph{AstroSat} and \emph{Swift}-\emph{Nu}STAR observations. This result agrees with the previous ones shown by \cite{Acciari2010b} and \cite{Costa2018} using UVOT observations. \cite{Costa2018} demonstrated this prominent extra thermal component along with optical flux points and concluded that the optical/UV emission is a separate component from a different emission origin. The work has the complete illustration of the SED modelling, using a giant elliptical galaxy template from \cite{Silva1998} to resolve the discrepancy seen in optical/UV regime. The far-IR emission from WISE data seems to be consistent with being due to the host galaxy cotribution, which may belong to the same spectral component as optical/UV. We see a marginal difference in the UV fluxes between observations in 2016 with UVIT and 2015 with UVOT, by a factor $\sim$ 1.5. Note that X-ray fluxes vary significantly during the period while the UV counterpart is relatively steady (see Table \ref{uvit}). This is a further evidence that both X-ray and UV emissions belong to separate emission origins.

\section{Summary and Discussion}
\label{discussion}

The unprecedented true simultaneous data from \emph{AstroSat} in broad X-ray energy range are crucial to establish the exact nature of synchrotron spectrum of extreme high energy peaked BL Lac sources. Unfortunately, the X-ray spectral study in different flux states has not been reported due to the unavailability of simultaneous X-ray observations. With the advent of \emph{AstroSat}, we are able to perform such a comparative study. The major aspects of the present work through spectral analysis are, 1) validate the exact location of synchrotron spectral peak within the observational range, 2) to interpret the variation in the multi-epoch X-ray spectra, and 3) verify the peculiar behaviour seen in optical/UV emissions using \emph{AstroSat}-UVIT observations.  

We investigate the X-ray spectrum of the EHBL source RGB\,J0710+591 for two different epochs, observed by SXT-LAXPC in November 2016 and XRT-\emph{Nu}STAR in September 2015. The synchrotron peaks of these X-ray spectra are well confined at high energies within the observation range. The major outcome of this study is the observed sharp curvatures and the phenomenal change in the synchrotron peaks. The sharp curvature is better explained by the synchrotron spectrum 
associated with the decline of the underlying electron number density around the maximum attainable electron energy in the acceleration region. This enable us to estimate the maximum available electron energy $\gamma_{max}$ in the blazar emission region and the acceleration timescale in terms of the source magnetic 
field and the Doppler factor. Consistently, the observed X-ray spectral evolution can be understood as a result of varying magnetic field/Doppler factor or with the particle acceleration time. The other important result is the optical/UV emissions observed with \emph{AstroSat}-UVIT. These observations reconfirm the 
optical/UV spectrum corresponds to a different emission component other than X-ray.

An important constraint in estimating the maximum available electron energy and the associated acceleration time scale is the lack of knowledge related to the 
source magnetic field and the jet Doppler factor. In the present work, we have assumed these quantities to be same as the one estimated by \cite{Costa2018} 
through broadband spectral modelling during September 2015. However, its been understood that the magnetic field and the jet Doppler factor often seem to vary 
during different flux states. This discrepancy can be overcome with the simultaneous information available at $\gamma$-ray energies. Further, 
the estimated $\gamma_{max}$ can be cross checked through VHE observations. \\ \\


\textbf{Acknowledgements}: This research has used the data of \emph{\emph{AstroSat}} mission of the Indian Space Research Organisation (ISRO), archived at the Indian Space Science Data Centre (ISSDC). The authors would like to acknowledge the support from the LAXPC Payload Operation Center(POC) and SXT POC at the TIFR, Mumbai for providing support in data reduction. This work has been performed utilizing the calibration data-bases and auxillary analysis tools developed, maintained and distributed by \emph{AstroSat}-SXT team with members from various institutions in India and abroad. The work has also made use of data, software, and/or web tools obtained from NASAs High Energy Astrophysics  Science Archive Research Center (HEASARC), a service of the Goddard Space Flight Center and the Smithsonian Astrophysical Observatory. RM, RG and PG would like to thank the Indian Space Research organisation (ISRO), Department of Space, India for the grant under $"$Space Science Promotion$"$. SC acknowledges CSR-NWU Potchefstroom, SA-GAMMA and NRF-South Africa for supporting his AstroSat $\&$ HESS related projects. PG would like to thank IUCAA and BARC for their hospitality.


\bibliographystyle{mnras}
\bibliography{rgb} 

\bsp	
\label{lastpage}

\end{document}